\UseRawInputEncoding
\documentclass[superscriptaddress, notitlepage, reprint]{revtex4-1}
\usepackage[english]{babel}
\usepackage{amsmath,amsthm}
\usepackage{amsfonts}
\usepackage[pdfborder={0 0 0}, colorlinks=true, urlcolor=blue, linkcolor=blue, citecolor=blue]{hyperref}
\usepackage{color}
\usepackage{graphicx}
\usepackage{float}
\usepackage{subfigure}
\usepackage{lipsum}
\usepackage{epstopdf}
\usepackage{dcolumn}
\begin{document}

\title{Quantum estimation of tripartite coupling in Spin-Magnon-Mechanical Hybrid Systems}
\author{Dong  Xie}
\email{xiedong@mail.ustc.edu.cn}
\affiliation{College of Science, Guilin University of Aerospace Technology, Guilin, Guangxi 541004, People's Republic of China}
\author{Chunling Xu}
\affiliation{College of Science, Guilin University of Aerospace Technology, Guilin, Guangxi 541004, People's Republic of China}

\begin{abstract}
Tripartite interactions play a fundamental role in the quantum information processing and quantum technology. However, it is generally difficult to
realize strong tripartite coupling. We investigate the estimation of a tripartite coupling strength in a hybrid setup
composed of a single nitrogen-vacancy (NV) center and a micromagnet. A time-independent parametric drive can be utilized to increase the estimation precision of the tripartite coupling strength. By calculating the quantum Fisher information (QFI), we can obtain the optimal estimation precision by measuring the eigenstate of the tripartite system. At the critical position, the QFI is divergent due to that the preparation time of the eigenstate is divergent. When the system is subjected to a dissipation, the QFI near the critical point of the driven-dissipation phase transition is analytically obtained. The direct intensity measurement is the optimal measurement near the dissipation phase transition point. In addition, we quantify the robustness of an imperfect measurement operator by the measurement noise susceptibility based on the error propagation formula. We find that the direct intensity measurement is enough robust against  small measurement
disturbance from a coherent drive. But it can be disturbed by the nonlinear anti-harmonic measurement noise, especially near the critical point.
\end{abstract}
\maketitle

\section{Introduction}
In the field of quantum information processing, the coupling between quantum systems is an important foundation to realize various purposes~\cite{lab1,lab2}.
The pairwise coherent coupling between a two-level quantum system and a quantized field is described by the Jaynes-Cummings mode, which is a typical quantum regime in the quantum optics~\cite{lab3,lab4}. In order to perform more complex tasks, the exploration of interactions beyond the pairwise interactions is becoming increasingly important and appealing~\cite{lab5}. However, the realization and control of the tripartite interactions are more difficult than the pairwise interactions.

A lot of previous studies mostly focus on using pairwise interactions to construct hybrid quantum setups based on magnons in microscopic magnets~\cite{lab6,lab7,lab8,lab9,lab10,lab11,lab12,lab13}, mechanical motions~\cite{lab14,lab15,lab16,lab17}, and nitrogen-vacancy (NV) centers in diamond~\cite{lab18,lab19,lab20,lab21}. Recently, it is shown that the tripartite interaction among single spins, magnons, and phonons can be obtained in a hybrid setup comprising a single NV center in diamond and a micromagnet~\cite{lab22}. The tripartite coupling can be enhanced by a parametric drive to amplify the mechanical zeropoint fluctuations of the vibration mode~\cite{lab23}.

Quantum estimation mainly aims at exploiting quantum resources to increase the estimation precision of measurements~\cite{lab24}.
Previous studies focus on quantum estimation in the pairwise interaction system.  The critical point of quantum phase transition has been utilized to improve the quantum estimation precision in the quantum Rabi system~\cite{lab25,lab26}. How to use critical resources to improve the measurement precision of tripartite coupling strength is of great significance.

In this article, we investigate the quantum estimation of the tripartite coupling strength in a hybrid setup comprising a single NV center in diamond and a micromagnet.  A time-independent parametric drive is proposed to increase the estimation precision of the tripartite coupling strength.
The optimal estimation precision is achieved by measuring the eigenstate of the tripartite system. The relative position between the NV center and the micromagnet
can be the extra freedom, which is used to obtain the critical point.  At the critical position, the QFI is divergent due to that the preparation time of the eigenstate is divergent. When the system is subjected to a dissipation process, the QFI near the critical point of the driven-dissipation phase transition is analytically obtained. The direct intensity measurement is the optimal measurement near the critical point. The measurements are often imperfect, which will reduce the estimation precision.  We quantify the robustness of an imperfect measurement operator by the measurement noise susceptibility based on the error propagation formula. The direct intensity measurement is enough robust against  small measurement
disturbance from a coherent drive. However, it can be disturbed by the nonlinear anti-harmonic measurement noise, especially near the critical point.

This article is organized as follows. In Section II, we introduce the tripartite interaction Hamiltonian, which can be obtained in spin-magnon-mechanical hybrid systems, and the QFI based on the eigenstate is achieved. In Section III, the dissipation dynamic evolution is derived. In Section IV, the optimal estimation precision of the tripartite is obtained in the dissipation process. In Section V, the direct intensity measurement is shown to be the optimal measurement at the critical point. The feasibility is discussed in Subsection VI. The measurement noise susceptibility is proposed to quantify the robustness of measurement operator in section VII.

 \section{Spin-Magnon-Mechanical Hybrid Systems}
\begin{figure}[h]
\includegraphics[scale=0.6]{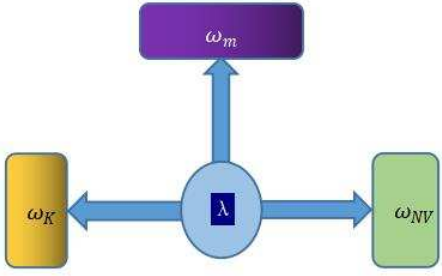}
 \caption{\label{fig.1}Schematic diagram of a spin-magnon-mechanical hybrid system. $\lambda$ represents the tripartite coupling strength among the NV center spin with the frequency $\omega_{NV}$, magnon with the frequency $\omega_K$, and mechnical mode with the frequency $\omega_m$. }
\end{figure}

We focus on the tripartite interaction Hamiltonian, which is described by ($\hbar=1$ through the whole article)
\begin{align}
H_{\textmd{Tri}}=\lambda (b+b^\dagger)(a+a^\dagger)\sigma_x,
\tag{1}
\end{align}
where $\lambda$ is the tripartite coupling strength to be estimated. The tripartite interaction Hamiltonian can appear in spin-magnon-mechanical hybrid systems composed of magnon [such as a yttrium iron garnet (YIG) sphere] and NV center in diamond, as shown in Fig.~\ref{fig.1}.

The free Hamiltonian  of the magnon is given  by the Kittel mode
\begin{align}
H_\textmd{K}=\omega_\textmd{K}a^\dagger a,
\tag{2}
\end{align}
where $\omega_\textmd{K}=\gamma B_Z$, the gyromagnetic ratio $\gamma$, a large external magnetic field $B_Z$.
The free Hamiltonian of NV center in diamond as a magnetic dipole is given by
\begin{align}
H_{\textmd{NV}}=\omega_{\textmd{NV}}\sigma_z/2,
\tag{3}
\end{align}
where the Pauli operators $\sigma_i$ are defined in the basis $\{|g\rangle,|e\rangle\}$ and $\omega_{\textmd{NV}}$ denotes the resonance frequency of the NV center.
The interaction Hamiltonian between the magnetic
field of the magnon and the NV center is described by
\begin{align}
H_{\textmd{int}}=-(g_e\mu_B)\hat{\vec{B}}(\vec{r})\cdot\hat{\vec{S}},
\tag{4}
\end{align}
where the Land\'{e} factor $g_e$, Bohr magneton $\mu_B$, and spin operators $\hat{\vec{S}}=1/2(\hat{\sigma}_x,\hat{\sigma}_y,\hat{\sigma}_z)$.

We consider that the magnetic field generated by the magnon (YIG) is along the coordinate vectors $\vec{e}_x$~\cite{lab22}
\begin{align}
\hat{\vec{B}}(\vec{r})=-\frac{\mu_0\sqrt{3 |\gamma|M_sR^3/(8\pi )}}{3 r^3}\vec{e}_x,
\tag{5}
\end{align}
where $R$ is a micromagnet of radius, $\mu_0$ is the permeability of vacuum, $M_s$ is the saturation magnetization, and $r$ is the distance between the NV center and the YIG sphere.
$r$ can be reexpressed as $r=r_0+z$ with $r_0$ denoting the equilibrium part of the distance.

Up to the first order of the coordinate $z$, the center-of-mass
vibration (phonon)  is quantized by $\hat{z}=z_{zpf}(\hat{b}+\hat{b}^\dagger)$ with with the zero-point fluctuation $z_{zpf}=\sqrt{1/(2M\omega_m)}$ to obtain the interaction Hamiltonian as
\begin{align}
H_{\textmd{int}}=\lambda (b+b^\dagger)(a+a^\dagger)\sigma_x+g_0(a+a^\dagger)\sigma_x
\tag{6}
\end{align}
with the pairwise coupling $g_0=\frac{\lambda r_0}{3z_{zpf}}$ and the tripartite spin-magnon-phonon coupling strength
\begin{align}
\lambda=\frac{3g_e\mu_0\mu_B}{8\pi r_0^4}\sqrt{\frac{4\pi|\gamma|M_sR^3}{3M\omega_m}},
\tag{7}
\end{align}
where $\omega_m$ denotes the vibration frequency of the phonon mode, and $M$ is the effective mass of the phonon mode.

Then, the total Hamiltonian of spin-magnon-mechanical hybrid system is described as
\begin{align}
H_{\textmd{Tot}}=&\omega_Ka^\dagger a+\omega_mb^\dagger b+\omega_{NV}\sigma_z/2\nonumber\\
&+\lambda (b+b^\dagger)(a+a^\dagger)\sigma_x+g_0(a+a^\dagger)\sigma_x
\tag{8}
\end{align}

In order to enhance the tripartite coupling strength, the the center-of-mass motion of the trapped diamond particle can  be driven by an additional  electrical
potential~\cite{lab27} with the time-independent stiffness Hamiltonian
\begin{align}
H_\textmd{d}=-\Omega_p({b^2}^\dagger+b^2)\tag{9}
\end{align}
In the squeezed frame by applying the unitary transformation $U_s(r)=\exp[r(b^2-{b^2}^\dagger)]$, the
total Hamiltonian can be written as

\begin{align}
H^s_{\textmd{Tot}}=&\omega_\textmd{K}a^\dagger a+\Delta_mb^\dagger b+\omega_{\textmd{NV}}\sigma_z/2+\nonumber\\
&\lambda_{e} (b+b^\dagger)(a+a^\dagger)\sigma_x+g_0(a+a^\dagger)\sigma_x
\tag{10}
\end{align}
where $\Delta_m=(\omega_m-\Omega_p)/\cosh 2r$ and $\lambda_{e}=\lambda e^r$ with the squeezing parameter $r$ defined as $\tanh 2r=\Omega_p/(\omega_m-\Omega_p)$.

Applying the Schrieffer-Wolff transformation
\begin{align}
U=\exp[i\frac{1}{\omega_{\textmd{NV}}}[\lambda_e(b+b^\dagger)+g_0](a+a^\dagger)\sigma_y]
\tag{11}
\end{align}
under the limit condition $\omega_\textmd{K}/\omega_{\textmd{NV}}\rightarrow0$~\cite{lab28}, we can achieve the decoupling Hamiltonian
\begin{align}
H_{\textmd{Tot}}^S=\omega_\textmd{K}a^\dagger a+\Delta_mb^\dagger b+\omega_{\textmd{NV}}\sigma_z/2+\nonumber\\
\frac{1}{\omega_{NV}}[\lambda_{e} (b+b^\dagger)+g_0]^2(a+a^\dagger)^2\sigma_z.
\tag{12}
\end{align}

For a large squeezing parameter $r$,  $\Delta_m$ is close to 0. The $n$th eigenstate of the system is described as
\begin{align}
|\psi_n(\xi)\rangle=S(\xi)|n\rangle|\downarrow\rangle|x_b\rangle\tag{13}\label{13},
\end{align}
where the squeeze operator $S(\xi)=\exp\{(\xi/2){a^2}^\dagger-(\xi^*/2){a^2}\}$ with the squeezing parameter $\xi=-\frac{1}{4}\ln\{1-\frac{4\Lambda^2}{\omega_{\textmd{NV}}\omega_\textmd{K}}\}$
with $\Lambda=\lambda_ex_b+g_0$.

When $\frac{4\Lambda^2}{\omega_{\textmd{NV}}\omega_\textmd{K}}=1$, the squeezing parameter $\xi$ is divergent. It denotes that the transition between normal phase and the supperradiance phase appears. Then, we can obtain the critical position $x_\pm=\pm\frac{\sqrt{\omega_{\textmd{NV}}\omega_\textmd{K}}}{2}-g_0$. When $x_-<x_b<x_+$, the system is in the normal phase;  When $x_b<x_-$ or $x_b>x_+$, the system is in the supperradiance phase~\cite{lab29}.

Our goal is to estimate the strength of the tripartite coupling $\lambda$. The measurement precision of
$\lambda$ can be derived through the quantum Cram\'{e}r-Rao bound~\cite{lab30}
 \begin{align}
\delta \lambda\geq1/\sqrt{\mathcal{F}_\lambda},\tag{14}\label{14}
\end{align}
where $\mathcal{F}_\lambda$ denotes the QFI associated with the parameter $\lambda$.

For the pure state in Eq.~(\ref{13}), the QFI $\mathcal{F}_\lambda$ can be calculated by using the derivative of the wave function with
respect to the unknown parameter
 \begin{align}
\mathcal{F}_\lambda=4[\langle\partial_\lambda\psi_n(\xi)|\partial_\lambda\psi_n(\xi)\rangle-|\langle\psi_n(\xi)|\partial_\lambda\psi_n(\xi)\rangle|^2]\tag{15}\\
=\frac{\Lambda^2e^{2r}x_b^2(n^2+n+1)}{2(\frac{\omega_{NV}\omega_K}{4}-\Lambda^2)^2}\tag{16}
\end{align}

For large number of the magnons $n\gg1$, the measurement precision with Heisenberg scaling with respect to the number of magnons can be achieved, i.e., $\mathcal{F}_\lambda\propto n^2$.

At the critical point $\frac{\omega_{\textmd{NV}}\omega_\textmd{K}}{4}=\Lambda_c^2$, the QFI $\mathcal{F}_\lambda$ is divergent. The reason for this is that the preparation time of the eigenstate will diverge. In order to stay in the eigenstate, we assume that the coupling strength was adiabatically ramped from $\frac{\Lambda}{\sqrt{\omega_{\textmd{NV}}\omega_\textmd{K}}/2}=0$ to $\frac{\Lambda}{\sqrt{\omega_{\textmd{NV}}\omega_\textmd{K}}/2}<1$. The time of such an adiabatic sweep is given by
 \begin{align}
T\approx \left(2\gamma \omega_\textmd{K}\sqrt{1-{4\Lambda^2}/{\omega_{\textmd{NV}}\omega_\textmd{K}}}\right)^{-1}\tag{17}\label{17},
\end{align}
where $\gamma\ll1$ for satisfying the adiabatic condition\cite{lab30a}.
Using this time, the QFI $\mathcal{F}_\lambda$ can be repressed as
\begin{align}
\mathcal{F}_\lambda=\frac{8(\gamma\omega_KT)^4\Lambda^2e^{2r}x_b^2n^2}{\Lambda_c^4}\tag{18}\label{18}.
\end{align}
It shows that $\mathcal{F}_\lambda $ is proportional to the fourth power of time $T^4$ without any dissipation process.

\section{dissipation dynamic}
We consider that the total systems are suffering from dissipation process and coherent drive. The master equation can be described by
\begin{align}
\frac{d\rho}{dt}=&-i[H^S_{\textmd{Tot}}+H_{\textmd{cd}},\rho]+(\kappa_aD[a]+\kappa_aD[b]\nonumber\\
&+\kappa_{\sigma^-}D[\sigma^-])\rho,\tag{20}\label{20}
\end{align}
where the coherent driving Hamiltonian is $H_{\textmd{cd}}=iF(a+a^\dagger)$ with the driving strength $F$ and the dissipation superoperator is given by $D[c]\rho=2c\rho c^\dagger-c^\dagger c\rho-\rho c^\dagger c$ with $c=(a, b, \sigma^-)$.
The Langevin equations corresponding to Eq.~(\ref{20}) can be derived by the formula~\cite{lab31,lab32}

\begin{align}
\frac{d\mathcal{O}}{dt}=&i[H^S_{\textmd{Tot}}+H_{cd},\mathcal{O}]+\sum_{c=a, b, \sigma^-}-[\mathcal{O},c^\dagger](\kappa_cc-\sqrt{2\kappa_c}c_{in})\nonumber\\
&+(\kappa_cc^\dagger-\sqrt{2\kappa_c}c^\dagger_{in}[\mathcal{O},c],\tag{21}\label{21}
\end{align}
where $\mathcal{O}$ can be any operator and the input quantum Gaussian noise operators $c_{in}(t)$ at zero temperature are characterized as
\begin{align}
&\langle c_{in}(t) \rangle=\langle c^\dagger_{in}(t') \rangle=0,\ \ \langle c^\dagger_{in}(t) c_{in}(t') \rangle=0,\tag{22}\label{22}\\
&\langle c_{in}(t)c^\dagger_{in}(t') \rangle=\delta(t-t').\tag{23}\label{23}
\end{align}

Utilizing above equations, we can obtain the evolution dynamic according to Eq.~(\ref{21})
\begin{align}
\dot{X}_a&=-\kappa_aX_a+\omega_KP_a+A_{a+},\tag{23}\label{23}\\
\dot{P}_a&=-\kappa_aP_a-\omega_KX_a-2\sigma_x(\lambda e^rX_b+g)-F+A_{a-},\tag{24}\label{24}\\
\dot{X}_b&=-\kappa_bX_b+\omega_mP_b+A_{b+},\tag{25}\label{25}\\
\dot{P}_b&=-\kappa_bP_b-\omega_2X_b-2\sigma_x\lambda e^rX_a+A_{b-},\tag{26}\label{26}\\
\dot{\sigma_x}&=-\kappa_{\sigma^-}\sigma_x-\omega_z\sigma_y-2\sigma_zA_{\sigma^-+},\tag{27}\label{27}\\
\dot{\sigma_y}&=-\kappa_{\sigma^-}\sigma_y+\omega_z\sigma_x-2\sigma_zX_a(\lambda e^rX_b+g)+2\sigma_zA_{\sigma^--},\tag{28}\label{28}\\
\dot{\sigma_z}&=-2\kappa_{\sigma^{-}}\sigma_z-\kappa_{\sigma^{-}}+2\sigma_yX_a(\lambda e^rX_b+g)\nonumber\\
&\ +\sqrt{2\kappa_{\sigma^{-}}}(\sigma^-\sigma^\dagger_{in}+\sigma^+\sigma_{in}),\tag{29}\label{29}\
\end{align}
where the operators are defined as $A_{c+}=\sqrt{2\kappa_c}(c_{in}+c^\dagger_{in})$ and $X_c=c+c^\dagger$ and $P_c=i(c^\dagger-c)$.

In order to obtain the steady values, let the expectation values of the above operators be equal to 0, i.e., $\langle \mathbf{M}\rangle=0$ with $\mathbf{M}=(X_a,P_a,X_b,P_b,\sigma_x,\sigma_y,\sigma_z)$. As a result, we can obtain the stable steady values
\begin{align}
&\langle X_a\rangle=\frac{F\omega_a}{k_a^2+\omega_a^2}, \langle P_a\rangle=\frac{F\kappa_a}{k_a^2+\omega_a^2},\tag{30}\\
&\langle \sigma_z\rangle=-1/2,
\langle X_b\rangle=\langle P_b\rangle=\langle \sigma_x\rangle=\langle \sigma_y\rangle=0.\tag{31}
\end{align}
Then, we can obtain the linear Langevin equation using the mean field approximation by expanding an arbitrary operator
$M$ in the form of $M=\langle M\rangle+\delta M$. After careful deduction, we can get
\begin{align}
\dot{\delta{X}_a}&=-\kappa_a\delta X_a+\omega_K\delta P_a+A_{a+},\tag{32}\\
\dot{\delta{P}_a}&=-\kappa_a\delta P_a-\omega_K \delta X_a-2g\delta\sigma_x+A_{a-},\tag{33}\\
\dot{\delta{X}_b}&=-\kappa_b\delta X_b+\omega_m\delta P_b+A_{b+},\tag{34}\\
\dot{\delta{P}_b}&=-\kappa_b\delta P_b-\omega_m\delta X_b-2\lambda e^r\langle X_a\rangle\delta\sigma_x+A_{b-},\tag{35}\\
\dot{\delta{\sigma_x}}&=-\kappa_{\sigma^-}\delta\sigma_x-\omega_z\delta\sigma_y+A_{\sigma^-+},\tag{36}\\
\dot{\delta{\sigma_y}}&=-\kappa_{\sigma^-}\delta\sigma_y+g\delta X_a+\lambda e^r\langle X_a\rangle\delta X_b+\omega_z\delta\sigma_x\nonumber\\
&\ \ -2g\langle X_a\rangle\delta\sigma_z-A_{\sigma^--},\tag{37}\\
\dot{\delta{\sigma_z}}&=-2\kappa_{\sigma^{-}}\delta\sigma_z+2g\langle X_a\rangle \delta\sigma_y,\tag{38}
\end{align}

We consider that $\omega_{NV}\gg\omega_K\gg\omega_m$ and $(\kappa_{\sigma^-},\ \kappa_a)\gg\kappa_b$. As a consequence, we can adiabatically eliminate the modes $\delta a$ and $\delta\sigma_x$, $\delta\sigma_y$, $\delta\sigma_z$. By  careful derivation, the mode $\delta\sigma_x$ can be obtained
\begin{align}
\delta \sigma_x\approx-\frac{\lambda \langle X_a\rangle}{\omega_{NV}}.\tag{39}
\end{align}

The reduced evolution dynamic of the modes $\delta X_b$ and $\delta P_b$ is described by
 \[
 \left(
\begin{array}{ll}
\dot{\delta X_b}\\
\dot{\delta P_b}\\
  \end{array}
\right )= \mathbf{V}\left(
\begin{array}{ll}
\delta X_b\\
\delta P_b\\
  \end{array}
\right ) +\\
 \left(
\begin{array}{ll}
A_{b+}\\
A_{b-}\\
  \end{array}
\right ),\tag{40}\]
where the evolution matrix $\mathbf{V}$ is given by
 \[
 \mathbf{V}=\left(
\begin{array}{ll}
 -\kappa_b,\ \ \omega_m\\
\ \omega,\ \ \  -\kappa_b\\
  \end{array}
\right ),\tag{41}\]
where the parameter $\omega=2\lambda^2\langle X_a\rangle^2/\omega_Z-\omega_m$. The eigenvalues of  $\mathbf{V}$ are
\begin{align}
E_\pm=-\kappa_m\pm\sqrt{\omega_m\omega}.\tag{42}
\end{align}
The characteristic time $\tau$ for the system to reach the steady state is obtained by satisfying $e^{\mathbf{V}\tau}=0$. Therefore, the characteristic time $\tau$ is expressed as
\begin{align}
\tau=1/(\kappa_m-\sqrt{\omega_m\omega}).\tag{43}
\end{align}

Assuming that the cavity field has reached the steady state after a long-time evolution, the solutions are described
\begin{align}
\delta X_b=\int_0^\infty e^{-\kappa_mt}[\cosh(\sqrt{\omega \omega_m}t)A_{b+}\nonumber\\
+\sqrt{\frac{\omega_m}{\omega}}\sinh(\sqrt{\omega \omega_m}t)A_{b-}],\tag{44}\\
\delta P_b=\int_0^\infty e^{-\kappa_mt}[\cosh(\sqrt{\omega \omega_m}t)A_{b-}\nonumber\\
+\sqrt{\frac{\omega}{\omega_m}}\sinh(\sqrt{\omega \omega_m}t)A_{b+}].\tag{45}
\end{align}
\section{The optimal estimation precision from the QFI}
The corresponding variance matrix $\mathcal{C}$ for the quadrature operators $q=(b+b^\dagger)/\sqrt{2}$ and $p=i(b^\dagger-b)/\sqrt{2}$ can be derived by the above equations
\begin{align}
&\mathcal{C}_{11}=\frac{1}{2}(\langle X_2^2\rangle-\langle X_2\rangle^2)=\frac{2\kappa_2^2-\omega\omega_m+\omega_m^2}{2\Delta},\tag{46}\\
&\mathcal{C}_{22}=\frac{1}{2}(\langle P_2^2\rangle-\langle P_2\rangle^2)=\frac{2\kappa_2^2-\omega\omega_m+\omega^2}{2\Delta},\tag{47}\\
&\mathcal{C}_{12}=\frac{1}{2}[\langle (X_2P_2+P_2X_2)/2\rangle-\langle X_2\rangle\langle P_2\rangle]=\frac{\kappa_b(\omega+\omega_m)}{2\Delta},\tag{48}
\end{align}
where the parameter $\Delta=\kappa_m^2-\omega\omega_m$.

Due to that the effective Hamiltonian $H_{eff}=i\mathbf{V}$ is quadratic, the steady state of the system is Gaussian~\cite{lab33}.
The QFI of Gaussian state is obtained by~\cite{lab26,lab34}
\begin{align}
\mathcal{F}_\lambda=\frac{1}{2(1+P_\lambda^2)}\textmd{Tr}[(\mathcal{C}^{-1}\mathcal{C}'_\lambda)^2]+2\frac{P_\lambda^{'2}}{1-P_\lambda^4}\nonumber\\
+\langle\mathbf{X}^\top\rangle'_\lambda\mathcal{C}^{-1}\langle\mathbf{X}\rangle'_\lambda,\tag{49}
\end{align}
where $P_\lambda=\frac{1}{2\sqrt{\textmd{Det}\mathcal{C}}}$ and $\bullet'_\lambda$ is the term by term derivative of $\bullet$ with respect to $\lambda$.

When $\Delta$ is close to 0, the QFI can be achieved analytically
\begin{align}
\mathcal{F}_\lambda=\frac{16\omega_m^2\lambda^2e^{4r}\langle X_a\rangle^4}{\omega_{NV}^2\Delta^2},\tag{50}\\
=\frac{16\omega_m\lambda^2\tau^2e^{4r}\langle X_a\rangle^4}{\omega\omega_{NV}^2}.\tag{51}
\end{align}
Here, we can see that the QFI is proportional to the square of the evolution time. In contrast, the QFI is proportional to the fourth power of the evolution time in the absence of dissipation. It shows that the dissipation reduces the QFI from the fourth to the second power of the evolution time.
From the quantum Cram\'{e}r-Rao bound, the measurement precision of $\lambda$ is given by
\begin{align}
\delta\lambda\geq\frac{\sqrt{\omega}\omega_{NV}}{4\sqrt{\omega_m}\lambda\tau e^{2r}\langle X_a\rangle^2}\tag{52},
\end{align}
\section{The direct intensity measurement}
The QFI gives the measurement precision obtained by the optimal measurement operator. However, the optimal measurement may not be easy to do in the experiment.

 Next, the we calculate the measurement precision of some feasible operators by using the error propagation formula
\begin{align}
\delta\lambda|_O=\frac{\Delta O}{|\langle\partial O/\partial \lambda\rangle|},\tag{53}
\end{align}
where $\Delta O=\sqrt{\langle O^2\rangle-\langle O\rangle^2}$ denotes the variance of the specific measurement operator $O$.
With a direct intensity measurement $n=b^\dagger b$, the measurement precision of $\lambda$ can be given by
\begin{align}
\delta\lambda|_{n}=\frac{\sqrt{(\mathcal{C}_{11}+\mathcal{C}_{22})^2-1}}{|\partial (\mathcal{C}_{11}+\mathcal{C}_{22})/\partial \lambda|},\tag{54}
\end{align}
When close to the critical point $\Delta\rightarrow0$, we can obtain the analytical precision obtained by the direct intensity measurement
\begin{align}
\delta\lambda|_{n}=\frac{\sqrt{\omega}\omega_{NV}}{4\sqrt{\omega_m}\lambda\tau e^{2r}\langle X_a\rangle^2}.\tag{55}
\end{align}
From this result, we can see that the direct intensity measurement can get the result from the optimal measurement. Therefore, the direct measurement is the optimal measurement near the critical point.

Another common measurement is the homodyne detection with the quadrature operator $e^{i\theta}b+e^{-i\theta}b^\dagger$. Due to that the expectation value of the quadrature operator is still equal to 0, the information for $\lambda$ can not be achieved by the homodyne detection.
\section{the measurement noise susceptibility}
In this section, we investigate whether the measurement scheme is robust against measurement imperfections. In general, small measurement disturbances are difficult to avoid completely. In order to quantify the influence of measurement noise, we propose the measurement noise susceptibility based on the error propagation formula
\begin{align}
\chi[P_M,N_M,\lambda]=\lim_{\epsilon\rightarrow0}\frac{1}{\epsilon}[1-\frac{\delta^2\lambda|_{P_M}}{\delta^2\lambda|_{(1-\epsilon)P_M+\epsilon N_M}}],\tag{56}\label{56}
\end{align}
where $P_M$ denotes the perfect measurement, $N_M$ denotes the noise measurement, $(1-\epsilon)P_M+\epsilon N_M$ denotes the practical measurement.
When the measurement noise does not have any effect on the measurement precision of $\lambda$, the measurement noise susceptibility $\chi[P_M,N_M,\lambda]=0$.
Based on the measurement noise susceptibility, we try to investigate the robustness of the direct intensity measurement $P_d=a^\dagger a$.

When the noise measurement is a coherent drive $N_M=b^{\dagger j}+b^j$ with $j=1,2...$, we can achieve that $\delta^2\lambda|_{(1-\epsilon)P_d+\epsilon N_M}=\frac{1}{\delta^2\lambda|_M+\epsilon^2/(1-\epsilon)^2\Delta^2N/|\langle\partial O/\partial \lambda\rangle|^2}$. Utilizing the Eq.~(\ref{56}), we can obtain the measurement noise susceptibility $\chi[P_d,N_M=b^{\dagger j}+b^j,\lambda]=0$. It shows that the direct intensity measurement is roust against the general coherent drive.

Then, we consider that the noise measurement is the anti-harmonic term $\xi(b^\dagger b)^2$.  In order to deal with the higher order terms, we utilize the decoupling relation~\cite{lab35,lab36}
\begin{align}
\langle ABC\rangle\approx&\langle AB\rangle\langle C\rangle+\langle A\rangle\langle BC\rangle+\langle AC\rangle\langle B\rangle\nonumber\\
&-2\langle A\rangle\langle B\rangle\langle C\rangle,\tag{57}\\
\langle ABCD\rangle\approx&\langle AB\rangle\langle CD\rangle+\langle AD\rangle\langle BC\rangle+\langle AC\rangle\langle BD\rangle\nonumber\\
&-2\langle A\rangle\langle B\rangle\langle C\rangle\langle D\rangle.\tag{58}
\end{align}
After an analytical derivation, we can obtain the measurement noise susceptibility
\begin{align}
\chi[P_d,N_M=\xi(b^\dagger b)^2,\lambda]&=\nonumber\\
&\xi\left(2+8\langle b^\dagger b\rangle-\frac{2\langle b^\dagger b\rangle^3+2\langle b^\dagger b\rangle^2}{\partial \langle b^\dagger b\rangle/\partial \lambda}\right).\tag{59}
\end{align}
In general, the measurement noise susceptibility $\chi[P_d,N_M=\xi(b^\dagger b)^2,\lambda]$ is not equal to 0. As it approaches the critical point, the measurement noise susceptibility also goes to infinity. It shows that the direct intensity measurement is is more susceptible to the interference of the nonlinear anharmonic noise measurement at the critical point. Therefore, the strength of noise measurement $\xi$ must be small for improving the measurement precision.

\section{parameter feasibility}
To realize the time-independent parametric drive, a proper electric potential can be obtained by the static Paul trap~\cite{lab37}.
 For practical considerations, the decay rate of Kittel mode is given by $\kappa_a\simeq 10\textmd{MHz}$~\cite{lab38}. The  the decay rate of the mechanical mode is $\kappa_b=1\textmd{Hz}$\cite{lab39}. The decay rate of the NV center spin is about $\kappa_{\sigma^-}\approx1\textmd{kHz}$~\cite{lab40}. Hence, it satisfies that
$(\kappa_{\sigma^-},\ \kappa_a)\gg\kappa_b$. The mechanical frequency is about $\omega_m\approx10\textmd{kHz}$~\cite{lab41}. The frequency of the NV center spin can be $\omega_{\textmd{NV}}\approx 10\textmd{GHz}$ and the frequency of the magnon is about $\omega_\textmd{K}\approx1\textmd{GHz}$. Therefore, it satisfies that $\omega_{NV}\gg\omega_K\gg\omega_m$.

\section{conclusion }
In this article, we study on the quantum estimation the tripartite coupling strength. At the critical position of the mechanical mode, the QFI is divergent due to the divergent adiabatic preparation time. In the dissipation process, a coherent single-particle drive is utilized to obtain the driven-dissipation phase transition. The QFI around the critical point is analytically achieved, which shows that the dissipation reduces the QFI from the fourth to the second power of the evolution time. The direct intensity measurement is the optimal measurement near the critical point. In addition, we quantify the robustness of an imperfect measurement operator by the measurement noise susceptibility based on the error propagation formula. The direct intensity measurement is enough robust against  small measurement disturbance from a coherent drive. However, it can be disturbed by the nonlinear anti-harmonic measurement noise, especially near the critical point. Hence, it is necessary to reduce the disturbance from the nonlinear anti-harmonic measurement noise in the experiment.

\section*{Acknowledgements}
This research was supported by the National Natural Science Foundation of China under Grant No. 62001134, Guangxi Natural Science Foundation under Grant No. 2020GXNSFAA159047.

\end{document}